\documentclass[trackchanges,twocolumn]{aastex701}

\usepackage{amsmath}	
\begin{document}
\title{Observational Evidence of Solar Spicules Associated with Microfilament Eruptions Using DKIST}

\author{Qifan Dong} 
\affiliation{Yunnan Observatories, Chinese Academy of Sciences, Kunming 650216, People's Republic of China}
\affiliation{University of Chinese Academy of Sciences, Beijing 100049, People's Republic of China}
\email{dongqifan@ynao.ac.cn}

\author[0000-0003-2891-6267]{Xiaoli Yan}
\affiliation{Yunnan Observatories, Chinese Academy of Sciences, Kunming 650216, People's Republic of China}
\affiliation{Yunnan Key Laboratory of Solar Physics and Space Science, Kunming 650216, People's Republic of China}
\email{yanxl@ynao.ac.cn}
\correspondingauthor{Xiaoli Yan}
\email{yanxl@ynao.ac.cn}

\author[0000-0002-6526-5363]{Zhike Xue}
\affiliation{Yunnan Observatories, Chinese Academy of Sciences, Kunming 650216, People's Republic of China}
\affiliation{Yunnan Key Laboratory of Solar Physics and Space Science, Kunming 650216, People's Republic of China}
\email{zkxue@ynao.ac.cn}

\author[0000-0003-0236-2243]{Liheng Yang}
\affiliation{Yunnan Observatories, Chinese Academy of Sciences, Kunming 650216, People's Republic of China}
\affiliation{Yunnan Key Laboratory of Solar Physics and Space Science, Kunming 650216, People's Republic of China}
\email{yangliheng@ynao.ac.cn}

\author[0000-0003-4393-9731]{Jincheng Wang}
\affiliation{Yunnan Observatories, Chinese Academy of Sciences, Kunming 650216, People's Republic of China}
\affiliation{Yunnan Key Laboratory of Solar Physics and Space Science, Kunming 650216, People's Republic of China}
\email{wangjincheng@ynao.ac.cn}

\author{Yadan Duan}
\affiliation{Yunnan Observatories, Chinese Academy of Sciences, Kunming 650216, People's Republic of China}
\affiliation{Yunnan Key Laboratory of Solar Physics and Space Science, Kunming 650216, People's Republic of China}
\email{duanyadan@ynao.ac.cn}

\author[0000-0002-9121-9686]{Zhe Xu}
\affiliation{Yunnan Observatories, Chinese Academy of Sciences, Kunming 650216, People's Republic of China}
\affiliation{Yunnan Key Laboratory of Solar Physics and Space Science, Kunming 650216, People's Republic of China}
\email{xuzhe6249@ynao.ac.cn}

\author{Yian Zhou}
\affiliation{Yunnan Observatories, Chinese Academy of Sciences, Kunming 650216, People's Republic of China}
\email{zhouyian@ynao.ac.cn}

\author{Xinsheng Zhang}
\affiliation{Yunnan Observatories, Chinese Academy of Sciences, Kunming 650216, People's Republic of China}
\email{zhangxinsheng@ynao.ac.cn}

\author[0009-0003-9377-1989]{Zongyin Wu}
\affiliation{Yunnan Observatories, Chinese Academy of Sciences, Kunming 650216, People's Republic of China}
\affiliation{University of Chinese Academy of Sciences, Beijing 100049, People's Republic of China}
\email{wuzongyin@ynao.ac.cn}

\author{Guotang Wu}
\affiliation{Yunnan Observatories, Chinese Academy of Sciences, Kunming 650216, People's Republic of China}
\affiliation{University of Chinese Academy of Sciences, Beijing 100049, People's Republic of China}
\email{wuguotang@ynao.ac.cn}


\begin{abstract}

The formation mechanism of spicules is fundamentally important for understanding mass and energy transport from the chromosphere into the corona. Recent studies suggested that spicules may be powered by microfilament eruptions. However, direct observational evidence remains limited due to insufficient spatial resolution. Using high-resolution H$\alpha$ broadband observations from the Visible Broadband Imager (VBI) onboard the Daniel K. Inouye Solar Telescope (DKIST), we identify 30 spicule events triggered by microfilament eruptions in a quiet Sun region near the solar disk center on 2023 August 29. The detected microfilaments have an average length of $0.93\pm0.46$ Mm and a minimum length of 0.17 Mm, substantially smaller than previously reported minifilaments. We identify two distinct morphological classes of ejecta: individual spicules associated with smaller microfilaments, and enhanced spicular activities associated with larger microfilaments. Moreover, some events exhibit apparent twisting motions. All these high-resolution observations provide compelling evidence that spicules can be triggered by microfilament eruptions.

\end{abstract}


\section{Introduction} \label{sec:intro}

As early as 1875, \citet{Secchi+1875} observed small-scale jets above the chromosphere. This phenomenon has been continuously observed and studied for over a century, and it is referred to as ``spicules''. Spicules are ubiquitous, dynamic, small-scale jets in the solar chromosphere, manifesting as plasma protrusions ejected from the magnetic network. Due to their vast numbers and persistent activity, it has long been hypothesized that they serve as an important conduit for transporting mass and energy into the solar corona, potentially playing a crucial role in addressing the long-standing puzzles of coronal heating and solar wind acceleration \citep{DePontieu+etal+2007Science,Tian+etal+2014,Huang+etal+2019,Bate+etal+2022}.

Based on observations from the Hinode Solar Optical Telescope (SOT) in the Ca II H passband, \citet{DePontieu+etal+2007pasj} classified spicules into two types according to their properties. Type I spicules have slower velocities and the rising and then descending motions, primarily located in plage and active regions. Type II spicules have higher velocities, appearing to ascend and fade (with subsequent studies showing they eventually recede in the higher atmosphere), and are observed in quiet Sun regions and coronal holes, as well as in plage regions \citep{DePontieu+etal+2007pasj,Rouppe+etal+2009,Pereira+etal+2012,Skogsrud+etal+2015}. The formation of spicules is thought to be driven by multiple mechanisms, including the rebound shocks \citep{Hollweg+1982}, magnetic reconnection \citep{Gonzalez+etal+2017,Samanta+etal+2019}, Alfv\'en waves \citep{Kudoh+Shibata+1999,Iijima+2017} and vortex motions \citep{Snow+etal+2018,Qi+etal+2025}, and so forth (for detailed reviews, see \citealp{Sterling+2000,Sterling+2021,Skirvin+etal+2023}). The formation mechanism of Type I spicules is generally attributed to shock waves driven by the leakage of photospheric p-mode oscillations \citep{DePontieu+etal+2004,Hansteen+etal+2006}. The generation of Type II spicules is often associated with magnetic reconnection, particularly the magnetic activity occurring at the footpoints of magnetic flux tubes. 

Recent studies of large-scale solar eruptive events have provided new physical insights into the origin of spicules. Through systematic studies of coronal jets, \citet{Sterling+etal+2015Na,Sterling+etal+2016} found that such jets are generally associated with the eruption of minifilaments (MFs) (with an average scale of 8 Mm), and that the eruptions are linked to magnetic flux emergence and flux cancellation, which may indicate an underlying reconnection process. This ``minifilament eruptions (MFEs)'' has since gained broad support and has been further refined. Subsequent high-resolution and multiwavelength observations have further established the magnetic topology, thermodynamic evolution, and reconnection processes involved in MFEs, forming a coherent physical picture linking MFEs to coronal jet and macrospicule formation \citep{Duan+etal+2023,Li+etal+2023,Yang+etal+2023,Wang+etal+2024,Panesar+etal+2025,Huang+Wang+2025}. Within this context, while the formation of spicules is fundamentally driven by magnetic reconnection \citep{Samanta+etal+2019}, our observations suggest that some events involve processes associated with smaller-scale microfilament eruptions. In such cases, analogous to the specific subset of coronal jets modeled by \citet{Sterling+etal+2015Na}, the magnetic reconnection is closely linked to the MFE.

Building upon the hierarchy of solar eruptions, \citet{Sterling+Moore+2016} proposed that spicules might be scaled down versions of coronal jets, driven by even smaller microfilaments (with scales $\lesssim 10^3$ km). They discovered that the numbers of eruptions associated with filaments, minifilaments, and the hypothesized microfilaments, as well as the sizes of these filament structures, follow a consistent power-law distribution, which may imply a common physical mechanism operating across vastly different spatial scales. This hypothesis gained observational support from \citet{Sterling+etal+2020}, who analyzed  H$\alpha$ data of the Goode Solar Telescope (GST) \citep{Samanta+etal+2019} and identified ``enhanced spicular activities" preceded by microfilament-like features at their footpoints. These spicules exhibited twisting motions similar to those observed in coronal jets, providing empirical support for the mechanism of the microfilament eruption \citep{Duan+etal+2023}.

Definitive confirmation of the microfilament eruption triggering spicules has been limited by the resolution constraints of previous solar telescopes. The Daniel K. Inouye Solar Telescope (DKIST; \citealp{Rimmele+etal+2020}), with its unprecedented 4-meter aperture, provides a leap in spatial resolution. In particular, the Visible Broadband Imager (VBI; \citealp{Woger+etal+2021}) achieves diffraction-limited imaging of the photosphere and chromosphere, enabling us to probe spatial scales approaching those expected for microfilaments. In this study, we analyzed DKIST VBI data to investigate potential microfilaments at the footpoints of spicules with unprecedented spatial resolution. Section \ref{sec:OaM} details the DKIST data parameters and the data processing methods. Section \ref{sec:results} presents the observational results. Section \ref{sec:dis} provides the discussion and summary of our findings.

\begin{figure*}[ht!]
	\includegraphics[scale=1.0]{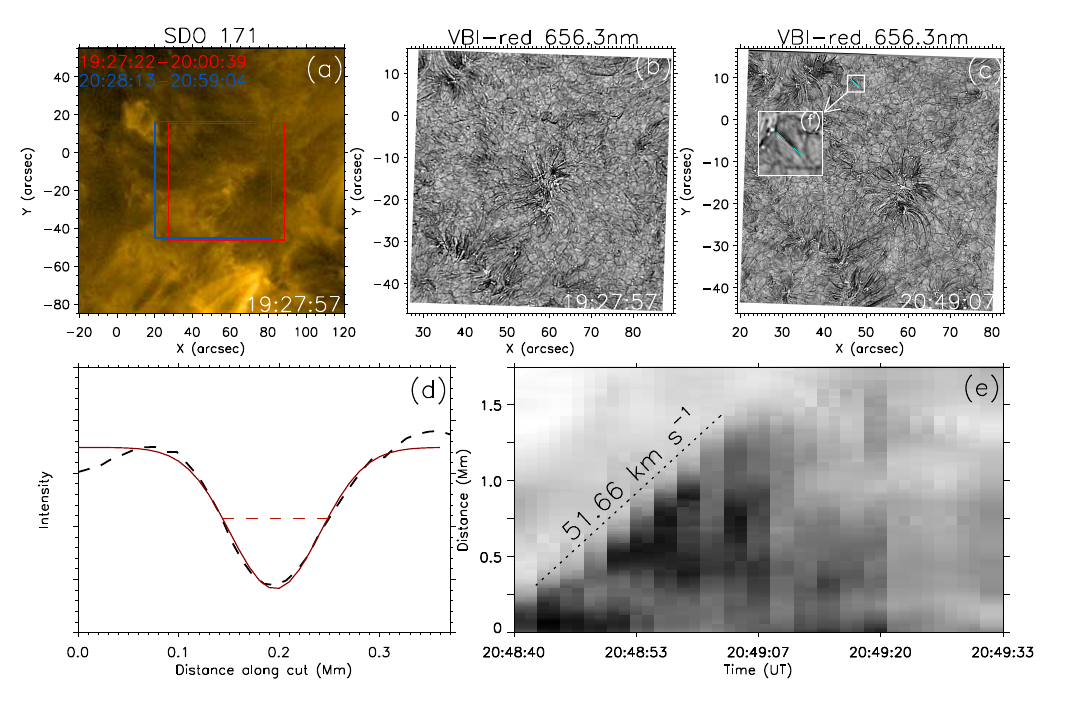}
	\caption{Overview of the FOV and methodology for statistical analysis. Panel (a) shows a large-FOV $171\,\mathrm{\AA}$ image. The red box and corresponding time interval denote the FOV of panel (b), while the blue box and corresponding time interval denote the FOV of panel (c). Panels (b-c) present H$\alpha$ broadband images observed by DKIST/VBI-red. Panel (d) displays the intensity profile used to derive the width of a microfilament; the black dotted line represents the observed data, the red solid line shows a single Gaussian fit, and the red dashed line marks the position of the FWHM. Panel (e) displays a space–time diagram obtained from a slice whose position in panel (c) is marked by the cyan line in the white box, and panel (f) shows a zoom‑in of the white box. The black dotted line is utilized to estimate the plane-of-sky velocity.}
	\label{fig:fig1}
\end{figure*}

\section{Observation and Method} \label{sec:OaM}
\subsection{Observation} \label{subsec:Ob}

The primary data used in this study were obtained from the VBI of the 4 m DKIST located in Hawaii. The VBI comprises two independent imaging channels, a blue channel (380–550 nm) and a red channel (550–850 nm), covering a total of eight spectral bands \citep{Woger+etal+2021}. As the primary objective of this study is to investigate dark, filament-like features located near and roughly perpendicular to spicule footpoints, we focus exclusively on H$\alpha$ observations from the VBI red channel and refer to these features as microfilaments hereafter. The H$\alpha$ band employs broadband imaging centered at 656.289 nm with a full width at half maximum (FWHM) of 0.047 nm.

To minimize the impact of projection effects on the measurement of microfilament properties, we selected data from a quiet Sun region close to solar disk center. By querying the DKIST Data Center Archive, we identified two sequences under Proposal ID pid\_2\_84, referred to as Data 1 (Product ID: L1-MQHAD) and Data 2 (Product ID: L1-TDWAC). Both datasets had a field of view (FOV) of $69^{\prime\prime} \times 69^{\prime\prime}$ with a spatial sampling of 0\farcs017 pixel$^{-1}$. The VBI acquired images at a cadence of 30 Hz. Speckle reconstruction was applied to every 80 short-exposure frames to obtain images approaching the diffraction limit of DKIST, resulting in an effective temporal cadence of 2.667 seconds for the reconstructed data \citep{Woger+etal+2008,Woger+etal+2021}.

The two datasets cover nearly identical FOVs but were acquired at different times (Figures \ref{fig:fig1}(a-c)). Data 1 was centered at coordinates (57\farcs92, -15\farcs54) and were observed from 19:27 UT to 20:00 UT on 2023 August 29, with an average Fried parameter (r\textsubscript{0}) of 10.4 cm. Data 2 was centered at (50\farcs92, -14\farcs54) and were observed from 20:28 UT to 20:59 UT, during which the r\textsubscript{0} was 11.7 cm. The image sequences were self-aligned using a cross-correlation method to remove image jitter.

To analyze the magnetic environment at microfilament locations and its potential impact on the upper atmosphere, data from the Solar Dynamics Observatory (SDO) were additionally utilized. Photospheric magnetic field information was obtained from the Helioseismic and Magnetic Imager (HMI) line-of-sight (LOS) magnetograms with a cadence of 45 s and a spatial sampling of 0\farcs5 pixel$^{-1}$, and a random noise level of approximately 10.2 G. Coronal responses were monitored using $171\,\mathrm{\AA}$ images from the Atmospheric Imaging Assembly (AIA), which have a temporal cadence of 12 s and a spatial sampling of 0\farcs6 pixel$^{-1}$.

\subsection{Method} \label{subsec:method}

Measurements of microfilament geometric and temporal parameters. The length and duration of microfilaments were extracted through manual identification and tracking. The length was defined as the maximum extended length reached during the entire evolution of the microfilament. The duration was defined as the time interval from its first appearance to its complete disappearance in the H$\alpha$ broadband images. The width of a microfilament was determined by performing a Gaussian fit to its transverse intensity profile. The intensity profile was extracted by taking a slice perpendicular to the axial direction of the microfilament when it reacheded its maximum length. The width ($w$) was taken as the FWHM of the fit, calculated as $ w = 2\sigma\sqrt{2\ln 2} $ (Figure \ref{fig:fig1}(d)), where $\sigma$ is the standard deviation of the Gaussian function.

Estimation of transversal velocity. The transversal velocity in the plane of the sky was estimated by measuring the transverse displacement of a single spicule during a given time interval. It is important to note that such transverse motions observed in 2D projections can originate from various physical processes. These primarily include: (1) the transfer of filament helicity via magnetic reconnection \citep{Sterling+etal+2020}; (2) magnetohydrodynamic (MHD) kink waves \citep{Kuridze+etal+2012}; (3) Alfvén waves \citep{DePontieu+etal+2007Science}; and (4) Kelvin–Helmholtz instabilities (KHI) \citep{Antolin+etal+2018}, which can generate transverse like motions through the development of vortex rolls at velocity shear interfaces, among others. The spicule positions were identified in two frames before and after the displacement. The average minimum Euclidean distance between the two positions was then calculated as the displacement distance, from which the transversal velocity in the plane of the sky was derived.

Measurement of plane-of-sky ejection velocity of spicule. The velocities of spicules were measured as the plane-of-sky velocity. This velocity was derived from space–time plots constructed along the direction of motion of the spicule. During the lifetime of the spicule, a slit was placed along its trajectory, and the intensity distribution along this path was extracted via interpolation at each time step. As the spicule exhibited approximately linear trajectories in the space–time plots, its plane-of-sky velocity was obtained by linearly fitting the slope of this trajectory (Figure \ref{fig:fig1}(e)).

\begin{figure*}[ht!]
	\includegraphics[scale=1.0]{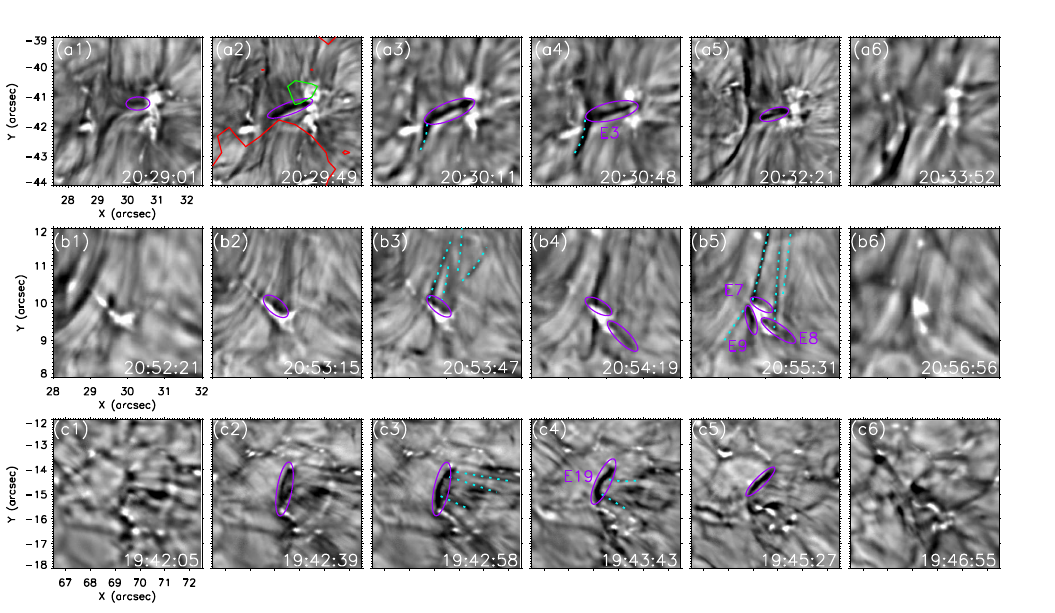}
	\caption{Evolution of five microfilament eruptions. The locations of the microfilaments are marked by purple ellipses, and the spicules produced by the eruptions are denoted by cyan dotted lines. In panel (a2), the red contour represents $+12$ G and the green contour represents $-12$ G.}
	\label{fig:fig2}
\end{figure*}

\section{Results} \label{sec:results}

We conducted a statistical analysis of the two datasets and identified a total of 30 spicule events. The properties of the microfilaments associated with these events are summarized in Table \ref{tab:Properties}, including length, width, duration, the presence or absence of transversal motion, the transversal velocity of the spicule, and the ejection velocity of the spicule. In the ``Transversal Motion'' column, ``Y'' denotes the presence of transversal motion and ``N'' its absence. It should be noted that the ejection velocities of spicules could not be reliably determined for all events, as in some cases the intensity contrast between the spicules and the background is insufficient for robust tracking in space–time plots. The average length of the microfilaments is $0.93\pm0.46$ Mm, which is slightly greater than the 0\farcs5–1\farcs0 range anticipated by \citet{Sterling+Moore+2016}, though at least half of the observed cases fall within this interval. Notably, three events have extremely small lengths with an average value of 0.24 Mm, and even with DKIST's high spatial resolution, these structures appeared nearly point-like. Interestingly, all three of these small-scale microfilaments were associated with individual spicules.

Figure \ref{fig:fig2} shows five representative events from three separate FOVs. In each row, the microfilaments are outlined by purple ellipses, while the spicules generated during the eruptions are marked by cyan dotted curves. Event 3 (Figures \ref{fig:fig2}(a1–a6)) is one of the few cases for which magnetic field information is available. Owing to the small spatial scales and weak magnetic fields of the microfilaments, usable SDO/HMI magnetograms are available for only three out of the 30 events. In Figure \ref{fig:fig2}(a2), the red and green contours indicate the LOS magnetic field levels of $\pm12$ G, slightly above the HMI random noise level (10.2 G), showing that the microfilament is located along a polarity inversion line (PIL). However, because the magnetic signal is close to the noise threshold, the measured field may be partially affected by instrumental noise. Furthermore, the spatial resolution of HMI (0\farcs5 pixel$^{-1}$) is insufficient to fully resolve the fine-scale magnetic structures associated with intergranular lanes at the scales relevant to this study, which may lead to spatial smearing and offset. The evolution of the microfilament is highlighted by the purple ellipse: it first appears in Figure \ref{fig:fig2}(a1), reaches its maximum length in Figure \ref{fig:fig2}(a3), and the cyan dotted curves in Figures \ref{fig:fig2}(a3–a4) trace the evolution of the associated spicule. Thereafter, the microfilament gradually fades and disappears by Figure \ref{fig:fig2}(a6). Events 7-9 (Figures \ref{fig:fig2}(b1–b6)) display three microfilaments clustered around a bright point, showing sequential evolution within a short time interval. The microfilament located above the bright point erupts first (Figure \ref{fig:fig2}(b2)). This microfilament has the longest duration (4.58 min) among the three and initially produces a cluster of spicules (Figure \ref{fig:fig2}(b3)). As the microfilament progressively disappears, the associated spicular activity evolves from a diffuse cluster into a single distinct spicule (Figure \ref{fig:fig2}(b5)). During this process, two additional microfilaments appear on both sides of the bright point and subsequently fade, accompanied by the formation of spicules (Figures \ref{fig:fig2}(b3–b5)). Event 19 (Figures \ref{fig:fig2}(c1–c6)) presents the evolution of another relatively long microfilament (1.52 Mm). During the evolution, a cluster of spicules (partly marked by cyan dotted lines in Figure \ref{fig:fig2}(c3–c4)) emerges, extending roughly perpendicular to the microfilament. Following this development, the microfilament gradually fades (Figure \ref{fig:fig2}(c5)), until both the microfilament and the associated spicules disappear by Figure \ref{fig:fig2}(c6). In contrast to the typical MFEs, no clear brightening is detected in the EUV passbands of AIA at the footpoint. However, the $171\,\mathrm{\AA}$ intensity integrated over the FOV of Figure \ref{fig:fig2}(a, b, c) peaks at the time of the eruption, suggesting that weak coronal heating may indeed be present, but it is too faint to be observed as a distinct brightening phenomenon. Such weak coronal signatures associated with chromospheric transients have been documented by \citet{Henriques+etal+2016}, and \citet{DePontieu+etal+2017} similarly reported coronal heating in association with spicules in both observations and simulations. In contrast to their findings, the size of our events is smaller, approaching or even smaller than the AIA resolution, which is why no obvious EUV brightenings were observed.

It should be noted that the proposal that enhanced spicular activities originate from microfilament eruptions \citep{Sterling+etal+2020} lacks direct observational confirmation. \citet{Sterling+etal+2020} found only tentative candidates, and \citet{Duan+etal+2023} studied larger scale macrospicules. Whether individual spicules share the same origin remains uncertain, and not all spicules are necessarily related to microfilaments \citep{Nobrega-Siverio+etal+2025}. The high spatiotemporal resolution of DKIST offers a way to test models.

\begin{figure}[ht!]
	\includegraphics[scale=0.65]{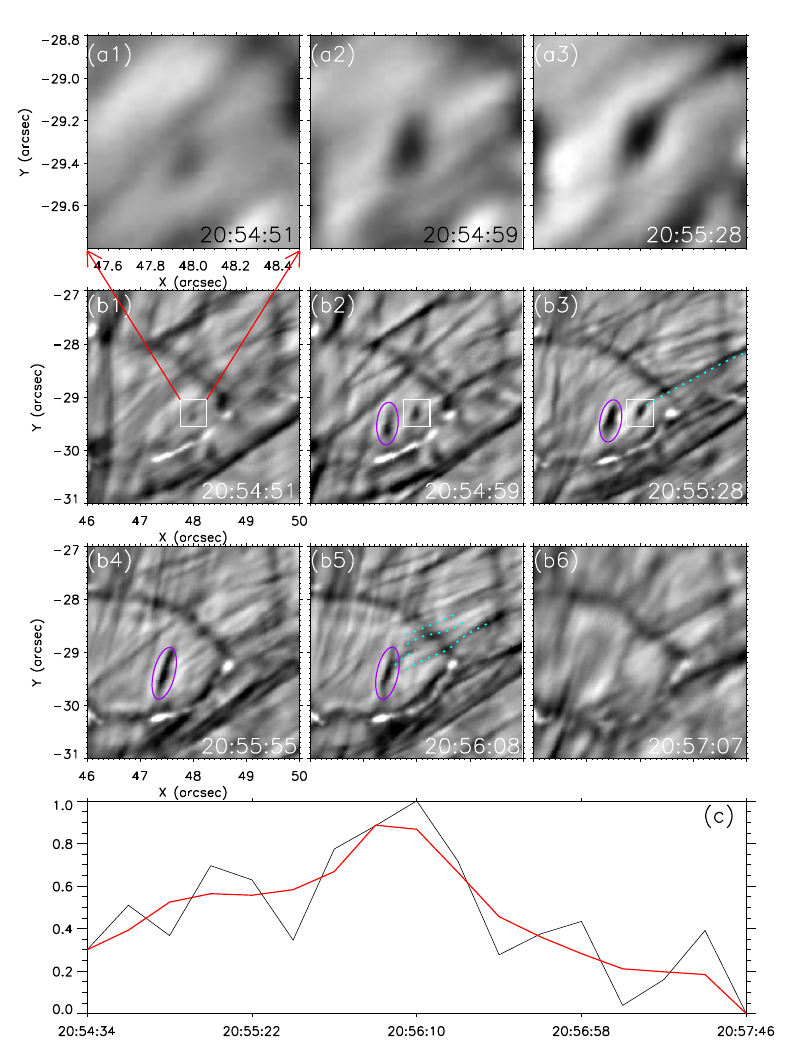}
	\caption{Comparison between an individual spicule and an enhanced spicular activity. Panels (a1–a3) show magnified views of the white-boxed region in panels (b1–b3), which depicts the evolution of a microfilament with a length of only 0.23 Mm; the cyan dotted line in panel (b3) marks the individual spicule generated by its eruption. Panels (b1–b6) illustrate the evolution of a larger-scale microfilament marked by a purple ellipse, with the cyan dotted line in panel (b5) indicating a cluster of spicules (enhanced spicular activity) produced by the eruption. Panel (c) presents the normalized $171\,\mathrm{\AA}$ intensity evolution over time within the region shown in panel (b1), where the black curve denotes the original data and the red curve represents the data smoothed over three time steps.}
	\label{fig:fig3}
\end{figure}

Figure \ref{fig:fig3} presents two microfilaments located in close proximity and evolving nearly simultaneously. The larger microfilament (Event 11), marked by a purple ellipse, has a length of 0.73 Mm and a duration of 2.44 min, its completion evolution (formation, development, and disappearance) is shown in Figures \ref{fig:fig3}(b1–b6). The smaller microfilament (Event 12), indicated by a white box, measures 0.23 Mm in length and lasts for 1.02 min, its evolution is displayed in Figures \ref{fig:fig3}(b1–b3), with enlarged views provided in Figures \ref{fig:fig3}(a1–a3). The uniqueness of these two cases lies in the fact that the smaller microfilament is associated with the formation of an individual spicule (cyan dotted line in Figure \ref{fig:fig3}(b3)), whereas the larger microfilament is associated with the formation of a cluster of spicules (cyan dotted lines in Figure \ref{fig:fig3}(b5)). This comparison may suggest that both individual spicules and enhanced spicular activities originate from microfilament eruptions, with the difference in spicule production likely arising from the spatial scale of the erupting microfilament. The black curve in Figure \ref{fig:fig3}(c) shows the $171\,\mathrm{\AA}$ intensity integrated over the FOV corresponding to Figure \ref{fig:fig3}(b1), with the red line representing a three time step smoothed version. The primary intensity peak occurs at 20:56:10 UT, coinciding with the eruption time of the larger microfilament, and a weaker peak near 20:55:10 UT corresponds to the eruption of the smaller microfilament. Although no obvious localized brightening was detected in the $171\,\mathrm{\AA}$ images, the temporal correspondence between the intensity enhancements and the eruption times may suggest that energy released by microfilament eruptions is transported into the quiet corona.

\begin{figure*}[ht!]
	\plotone{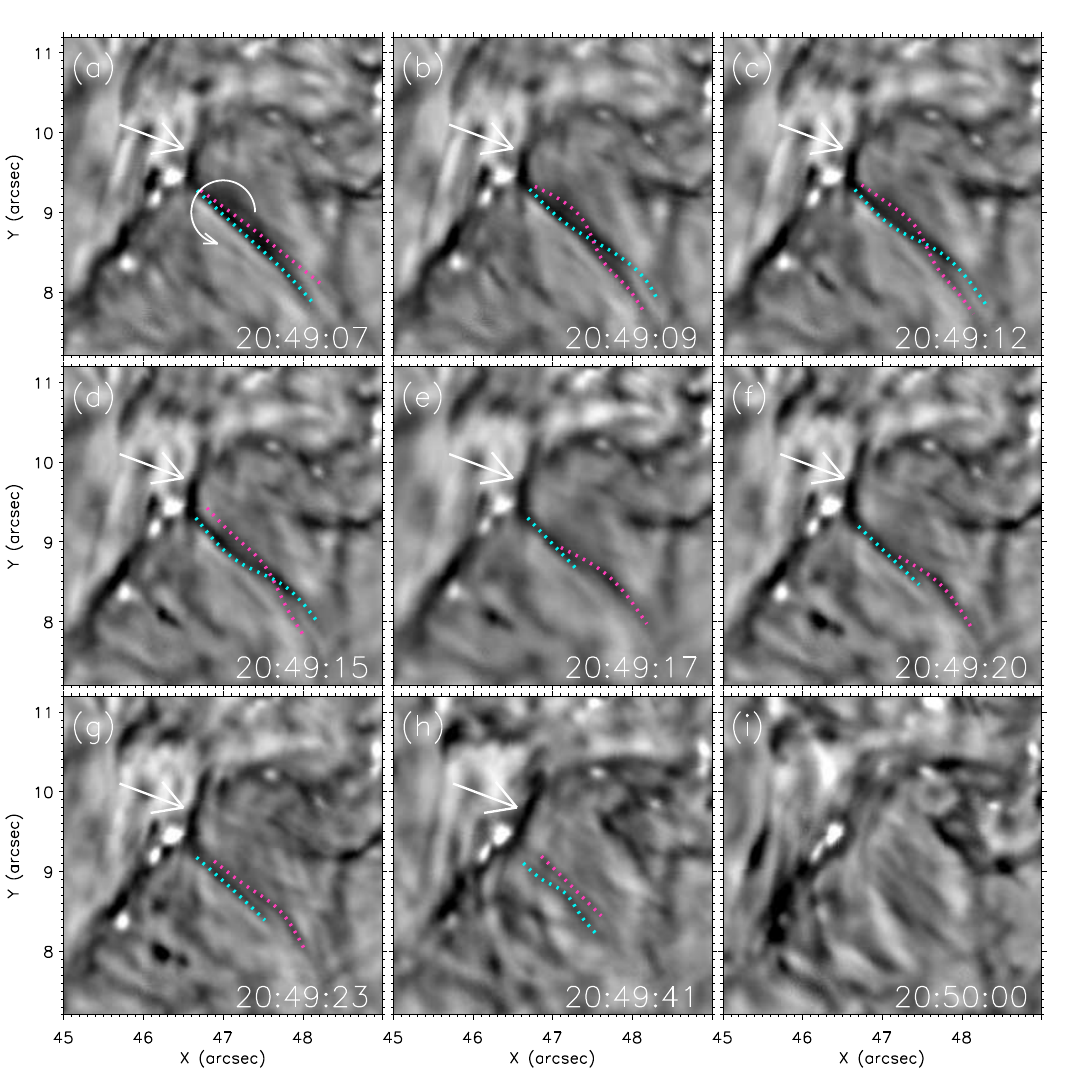}
	\caption{Transversal motions of spicules. The cyan and pink dotted lines trace two different spicules exhibiting rotational motions. The white arrow indicates the location of the associated microfilament. Curved arrows in panel (a) denote the direction of the spicule rotation.}
	\label{fig:fig4}
\end{figure*}

Since the VBI H$\alpha$ band employs broadband imaging without off-band or spectral data, Doppler velocities cannot be used to provide definitive evidence of transversal motion. Consequently, the presence of such motion was inferred by visually inspecting the apparent transverse/torsional motions of spicules in the image sequences. Among the 30 events, spicules in 11 events exhibited clear transversal motions, with an average velocity of $15.91 \pm 7.00\,\mathrm{km}\,\mathrm{s}^{-1}$. Figure \ref{fig:fig4} illustrates a representative example of spicule transversal motion. The cyan and pink dotted lines trace two spicules undergoing mutual transversal motion within the FOV, while the straight arrow indicates the location of the associated microfilament and the curved arrow marks the direction of the rotation. Figures \ref{fig:fig4}(a–h) clearly show the rotational evolution of the spicules. Shortly after the eruption, both the microfilament and the spicules fade, as shown in Figure \ref{fig:fig4}(i). This process may signify that the twisted magnetic field of the microfilament, through reconnection, transfers its magnetic helicity into the field lines of the spicules, thereby leading to the observed transversal motions. It should be emphasized that microfilament eruptions are only one of the possible physical mechanisms associated with spicule formation. As such, our observations show that spicules can still form even after the corresponding microfilament has completely dispersed.

\begin{figure*}[ht!]
	\plotone{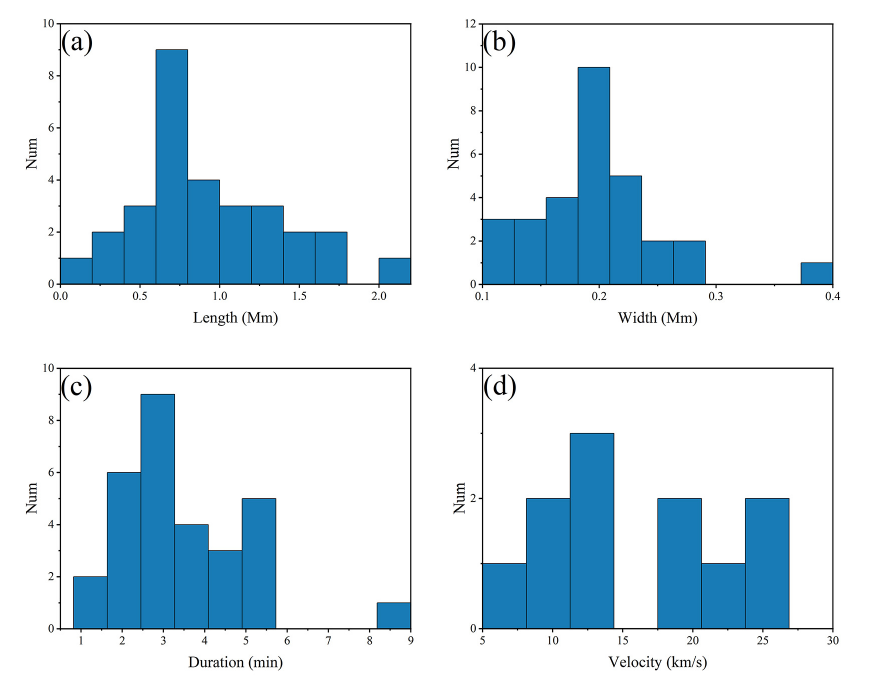}
	\caption{Histograms of microfilament properties including length (a), width (b), duration (c) and transversal velocity of enhanced specular activity (d).}
	\label{fig:fig5}
\end{figure*}

The statistical distributions of the physical parameters of the sampled microfilaments are presented in Figure \ref{fig:fig5}. The microfilaments exhibit a characteristic length (Figure \ref{fig:fig5}(a)) ranging from 0.2 to 2.0 Mm, with the majority concentrated between 0.5 and 1.5 Mm and a predominant population concentrated near 0.7 Mm. The width distribution (Figure \ref{fig:fig5}(b)) is dominated by a sharp peak around $\sim0.2\,\mathrm{Mm}$, with most measurements confined to 0.15–0.25 Mm. The duration distribution (Figure \ref{fig:fig5}(c)) features a peak at $\sim3\,\mathrm{min}$, with the bulk of durations falling between 1 and 6 min, and very few persisting beyond 7 min. Notably, this range closely resembles that reported in previous statistical studies of spicules. Finally, the kinematic distribution (Figure \ref{fig:fig5}(d)) shows apparent transversal velocities spanning $5-30\,\mathrm{km}\,\mathrm{s}^{-1}$. Owing to the limited number of events showing clear transversal motions (11 out of 30), the corresponding velocity distribution appears comparatively flat, with most values distributed between 10 and $25\,\mathrm{km}\,\mathrm{s}^{-1}$ and no pronounced peak.

\section{Discussion and Conclusion} \label{sec:dis}

In this study, microfilaments in the quiet Sun were investigated using H$\alpha$ broadband images acquired by the VBI red channel onboard DKIST. However, darkening in H$\alpha$ can arise from multiple structures and is not unique to microfilaments. In addition to filaments, H$\alpha$ dark features may correspond to fibrils \citep{Leenaarts+etal+2015}, mottles \citep{Kuridze+etal+2012}, spicules \citep{Pereira+etal+2012}, surges \citep{Jibben+2004}, blue and red shifted excursions (RBEs/RREs) \citep{Sekse+etal+2013} and other small-scale dynamic structures in the chromosphere. Furthermore, non-equilibrium effects following transient heating events can enhance H$\alpha$ opacity, producing dark fibril-like structures even in the absence of cool dense plasma \citep{Rutten+etal+2019}. Additionally, because these are H$\alpha$ broadband images, intergranular lanes may also affect the identification (see related discussion in Appendix \ref{sec:ruling}). Therefore, H$\alpha$ darkening alone does not uniquely diagnose as microfilament. In this work, our identification is based not solely on the presence of dark features, but on their temporal evolution, spatial morphology, magnetic field characteristics, and their close association with spicules. Within this framework, the microfilament eruption is interpreted as one possible pathway for spicule formation, rather than a unique mechanism. A total of 30 events were identified, nearly all of which were accompanied by the generation of spicules. A key distinction between our findings and previous observations lies in the fact that the filaments detected here exhibit markedly smaller spatial scales. The statistical results show that the microfilaments have an average length of $0.93\pm0.46$ Mm and a mean duration of $3.48\pm1.51$ min. These values are approximately an order of magnitude smaller than those of MFs, which typically range from 3 to 20 Mm in length with durations of less than 1 hour \citep{Wang+etal+2000}. Upon exclusion of the five events with lengths exceeding 1.5 Mm, the mean microfilament length decreases to $0.78\pm0.30$ Mm, which is in good agreement with the expected scale of 0\farcs5–1\farcs0. Furthermore, the observed microfilament durations match the spicule lifetime range (50–400 s), indicating a direct physical link between the two phenomena. The marked difference in spatial scale of microfilaments relative to MFs, along with their consistency in characteristics, provides observational evidence for the power-law distribution of solar eruptive features proposed by \citet{Sterling+Moore+2016}.

Numerous studies have established magnetic flux cancellation as the primary trigger of MFEs \citep{Panesar+etal+2017,Yang+etal+2024}. Given the extremely small spatial scales of the events, with some falling below the resolution of HMI, key magnetic parameters (e.g., field strength and configuration) are largely inaccessible. Nevertheless, three events in our sample possess valid magnetic field measurements (one of which is shown in Figure \ref{fig:fig2}(a2)), and all three are located along PILs. This suggests that magnetic flux cancellation may serve as an observational signature of the reconnection process involved in triggering microfilament eruptions at even smaller spatial scales.

One key finding of this study is that individual spicules and enhanced spicular activities share the same driving mechanism: microfilament eruptions. Despite the limited sample size, our results reveal that the morphology of the resulting spicules may be governed by the physical scale of the driving microfilament. Larger microfilaments ($\sim 0.7$ Mm) tend to drive enhanced spicular activities in the form of spicule clusters, in agreement with the results of \citet{Sterling+etal+2020}, whereas extremely small microfilaments ($\sim 0.2$ Mm) are connected with individual spicules. This relationship provides new insight, suggesting that the ``multiple threads'' appearance of spicules may fundamentally depend on the scale of the microfilament. When the erupting structure is at smaller scales ($< 0.4$ Mm), the eruption may manifest as a single, narrow plasma jet. Furthermore, the ejection velocities of individual spicules are substantial, reaching approximately $90\,\mathrm{km}\,\mathrm{s}^{-1}$. This may indicate that, for individual spicules, the magnetic energy released by extremely small-scale microfilaments is concentrated into a single magnetic structure, allowing more efficient conversion of magnetic energy into kinetic energy and thus producing higher ejection velocities.

The influence of physical scale on the dynamical properties is also reflected in the measured transversal velocities. Apparent transversal motions are detected in $37\%$ of the events, a behavior commonly attributed to the release and untwisting of magnetic helicity during magnetic reconnection. The measured transversal velocities are approximately $10-25\,\mathrm{km}\,\mathrm{s}^{-1}$. These values are lower than those reported for larger-scale eruptive phenomena, such as coronal jets ($20-50\,\mathrm{km}\,\mathrm{s}^{-1}$; \citealp{Sterling+etal+2015Na}) and enhanced spicular activities ($20-30\,\mathrm{km}\,\mathrm{s}^{-1}$; \citealp{Sterling+etal+2020}). This reduction in transversal velocity is consistent with the decreasing physical scale of the erupting structure, further supporting the hypothesis that solar eruptive events across a wide range of spatial scales are underpinned by a unified physical mechanism. 

\citet{Wang+etal+2024} estimated the energy of a $\sim 5^{\prime\prime} \times 2^{\prime\prime}$ MFE using high-resolution spectroscopic and magnetic field data from GST, and estimated its thermal and kinetic energies as $3.6 \times 10^{24}$ erg and $2.6 \times 10^{24}$ erg, respectively. The average scale of the microfilaments in this study is 1\farcs3 $\times$ 0\farcs2. Scaling the energy by the area ratio ($0.26/10$), a single microfilament eruption releases a thermal energy of $0.09 \times 10^{24}$ erg, a kinetic energy of $0.06 \times 10^{24}$ erg, and a total energy of $0.15 \times 10^{24}$ erg.

Over an observing interval of approximately one hour, a total of 30 microfilament eruptions were identified within a FOV of $\sim 65^{\prime\prime} \times 65^{\prime\prime}$ ($\sim 2227\,\mathrm{Mm}^{2}$). The event occurrence rate (R) is thus calculated as follows: 
\[
R = \!\frac{30}{1\,\mathrm{h} \times 2227\,\mathrm{Mm}^2} \approx 0.014\,\mathrm{events}\,\mathrm{hr}^{-1}\,\mathrm{Mm}^{-2}\!.
\]
Assuming that such events are uniformly distributed over the solar surface, with a total area of $\sim 6.1 \times 10^{6}$ Mm$^{2}$, the derived global microfilament eruption rate is about $8.5 \times 10^{4}$ events hr$^{-1}$. Combined with the average duration of microfilaments ($\sim 3.5$ min) obtained from the statistics, we estimate that roughly $2.4 \times 10^{4}$ microfilament eruptions exist globally at any given time. \citet{Sterling+etal+2016} previously estimated the global number of spicules to be in the range of $9.3 \times 10^{4}$ to $4.6 \times 10^{5}$. Although the estimated number of microfilaments is lower than the number of spicules, it is important to note that a microfilament is often associated with a cluster of spicules (as shown in Figure \ref{fig:fig3}). This order-of-magnitude consistency indicates that at least a part of spicules can be generated by microfilament eruptions.

In summary, using high spatial and temporal resolution H$\alpha$ observations from DKIST/VBI, we present direct observational evidence that spicules can be driven by microfilament eruptions. We have systematically analyzed 30 spicule events associated with microfilaments and their key characteristics. The most striking result is the identification of extremely small filaments with lengths down to 0.17 Mm, which, to our knowledge, represent among the smallest filaments reported in the solar atmosphere to date. Our results confirm that the scale of the microfilament likely determines the resulting spicular morphology, ranging from individual spicules produced by the small-scale microfilaments to enhanced spicular activities with pronounced twisting motions driven by larger microfilaments. These high-resolution observations provide a strong support for the idea that at least a fraction of spicules are driven by microfilament eruptions and suggest the existence of a unified eruptive mechanism operating continuously from coronal jets down to spicules, offering a new observational foundation for understanding mass and energy transport in the solar atmosphere.

\begin{deluxetable*}{lcccccc}
\tabletypesize{\scriptsize}
\tablewidth{0pt}
\setlength{\tabcolsep}{8pt} 
\renewcommand{\arraystretch}{1} 
\tablecaption{Properties of Microfilament Eruptions \label{tab:Properties}}
\tablehead{
\colhead{Event} & \colhead{Length\tablenotemark{a}} & \colhead{Width\tablenotemark{b}} & \colhead{Duration\tablenotemark{c}} & \colhead{Transversal Motion\tablenotemark{d}} & \colhead{Transversal Velocity\tablenotemark{e}} & \colhead{Ejection Velocity\tablenotemark{f}} \\
\colhead{} & \colhead{(Mm)} & \colhead{(Mm)} & \colhead{(min)} & \colhead{} & \colhead{(km\,s$^{-1}$)} & \colhead{(km\,s$^{-1}$)}
}
\startdata 
1             & 0.83 & 0.13 & 3.56 & N & \nodata           & 35.80 \\ 
2             & 0.99 & 0.14 & 5.47 & N & \nodata           & \nodata\\ 
3  (Figure \ref{fig:fig2}) & 1.02 & 0.16 & 5.20 & N & \nodata           & 51.32 \\ 
4             & 1.07 & 0.15 & 8.40 & Y & 19.34 $\pm$ 4.18  & \nodata\\ 
5             & 1.54 & 0.09 & 4.45 & N & \nodata           & \nodata \\ 
6             & 1.07 & 0.11 & 2.58 & N & \nodata           & \nodata\\ 
7  (Figure \ref{fig:fig2}) & 0.60 & 0.15 & 4.58 & Y & 22.93 $\pm$ 1.02  & 33.18 \\ 
8  (Figure \ref{fig:fig2}) & 0.78 & 0.13 & 2.80 & Y & 19.26 $\pm$ 1.13  & 132.52 \\ 
9  (Figure \ref{fig:fig2}) & 0.67 & 0.13 & 3.24 & N & \nodata           & \nodata\\ 
10            & 1.79 & 0.08 & 3.91 & N & \nodata           & \nodata\\ 
11 (Figure \ref{fig:fig3}) & 0.73 & 0.10 & 2.44 & Y & 12.27 $\pm$ 1.97  & \nodata\\ 
12 (Figure \ref{fig:fig3}) & 0.23 & 0.10 & 1.02 & N & \nodata           & 94.28 \\ 
13            & 2.12 & 0.12 & 5.29 & N & \nodata           & \nodata\\ 
14            & 0.86 & 0.08 & 2.09 & N & \nodata           & \nodata\\ 
15 (Figure \ref{fig:fig4}) & 0.78 & 0.13 & 2.31 & Y &  6.32 $\pm$ 2.09  & 51.66 \\ 
16            & 0.60 & 0.17 & 2.49 & N & \nodata           & \nodata\\ 
17            & 0.52 & 0.15 & 2.27 & N & \nodata           & 47.69 \\ 
18            & 0.17 & 0.10 & 2.09 & Y & 25.31 $\pm$ 3.34  & 88.27 \\ 
19 (Figure \ref{fig:fig2}) & 1.52 & 0.15 & 5.47 & Y & 11.73 $\pm$ 4.06  & 55.78  \\ 
20            & 1.31 & 0.19 & 2.36 & N & \nodata           & 34.67 \\ 
21            & 0.94 & 0.13 & 3.73 & N & \nodata           & \nodata\\ 
22            & 0.76 & 0.13 & 2.49 & N & \nodata           & 12.44 \\ 
23            & 1.62 & 0.19 & 4.27 & N & \nodata           & \nodata \\ 
24            & 0.74 & 0.13 & 3.56 & Y & 11.35 $\pm$ 0.61  & \nodata \\ 
25            & 0.32 & 0.27 & 1.60 & N & \nodata           & 106.98 \\ 
26            & 0.72 & 0.16 & 4.93 & Y & 10.53 $\pm$ 0.44  & 48.80 \\ 
27            & 1.31 & 0.14 & 2.71 & Y &  9.40 $\pm$ 3.79  & \nodata \\ 
28            & 0.52 & 0.18 & 3.16 & N & \nodata           & 20.47 \\ 
29            & 1.21 & 0.14 & 3.20 & Y & 26.60 $\pm$ 3.41  & 27.45 \\ 
30            & 0.68 & 0.14 & 2.89 & N & \nodata           & 26.97 \\ 
\hline
Averages      & 0.93 $\pm$ 0.46 & 0.14 $\pm$ 0.04 & 3.48 $\pm$ 1.51 & \nodata & 15.91 $\pm$ 7.00 & 54.27 $\pm$ 33.94 \\ 
\enddata
\tablenotetext{a}{Maximum length of the microfilament during its evolution.}
\tablenotetext{b}{Width of the microfilament.}
\tablenotetext{c}{Duration from microfilament appearance to its eruption and disappearance.}
\tablenotetext{d}{Transversal motions of spicules triggered by microfilament eruptions: ``Y'' = apparent spinning, ``N'' = no transversal motion.}
\tablenotetext{e}{Relative transversal velocity of the spicule.}
\tablenotetext{f}{Plane-of-sky ejection velocity of spicules generated by microfilament eruptions.}
\end{deluxetable*}

\begin{acknowledgments}
The research reported herein is based in part on data collected with the Daniel K. Inouye Solar Telescope (DKIST), a facility of the National Solar Observatory (NSO). NSO is managed by the Association of Universities for Research in Astronomy, Inc., and is funded by the National Science Foundation. Any opinions, findings and conclusions or recommendations expressed in this publication are those of the authors and do not necessarily reflect the views of the National Science Foundation or the Association of Universities for Research in Astronomy, Inc. In addition, the authors acknowledge the data support from SDO. This work is supported by the Strategic Priority Research Program of the Chinese Academy of Sciences, grant No. XDB0560000; the National Science Foundation of China (NSFC) under Nos. 12325303, 12473059, 12433010, and 12403065; Yunnan Key Laboratory of Solar Physics and Space Science under No. 202205AG070009; the Yunnan Science Foundation of China under Nos. 202501AW070002 and 202601AS070076; the Yunnan Province Xing Dian Talent Support Program, the Yunling Scholar Project, and Yunnan Revitalization Talent Support Program under Nos. 202305AS350029, 202305AT350005.
\end{acknowledgments}

\appendix

\section{High-Resolution Magnetic Field Evidence from DKIST/ViSP}
\setcounter{figure}{0}                
\renewcommand{\thefigure}{A\arabic{figure}}   

\begin{figure*}[ht!]
	\plotone{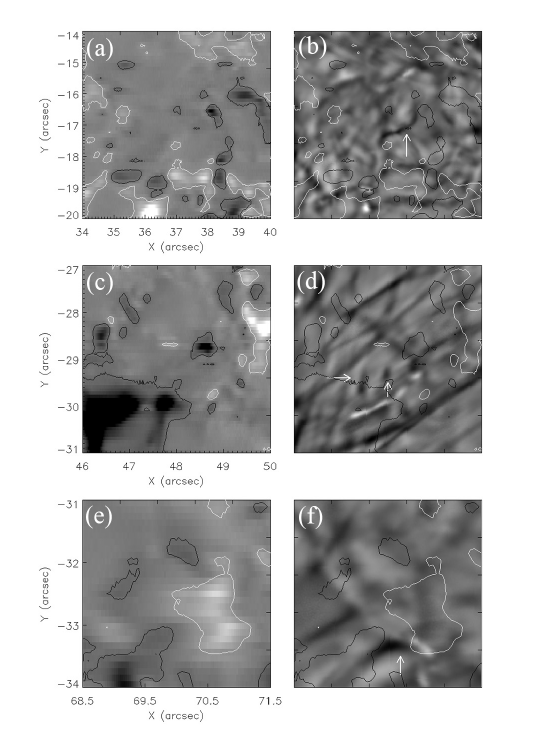}
	\caption{Panels (a), (c), and (e) show the high-resolution LOS magnetic field inferred from the circular polarization derived from ViSP Stokes I and V via the WFA. The white and black contours indicate positive and negative polarities, respectively, with contour levels at ±5 G. Panels (b), (d), and (f) present the corresponding VBI H$\alpha$ images. The white arrows mark the locations of the microfilaments.}
	\label{fig:figA1}
\end{figure*}

Characterizing the magnetic topology at the observed micro-scale requires a resolution beyond the capabilities of SDO/HMI. To better constrain the LOS magnetic field, we have incorporated high-resolution spectropolarimetric data from the DKIST Visible Spectro-Polarimeter (ViSP), targeting the Fe I 525.04 nm line. The high-resolution LOS magnetic field ($B_{\text{LOS}}$) shown in the Figures \ref{fig:figA1}(a, c, e) is derived using the weak-field approximation (WFA), which utilizes both the Stokes I and V profiles to compute the magnetic field, and the displayed quantity represents the magnetic flux density ($B_{\text{LOS}}$). As shown in Figure \ref{fig:figA1}, for the subset of events covered by ViSP, the dark features are preferentially located along or in close proximity to small-scale polarity inversion lines (PILs) or mixed-polarity regions. This spatial association is consistent with expectations for filament-like structures and provides supporting evidence for the interpretation as microfilaments.

\section{Ruling out intergranular lanes} \label{sec:ruling}
\setcounter{figure}{0}                
\renewcommand{\thefigure}{B\arabic{figure}}   

\begin{figure*}[ht!]
	\plotone{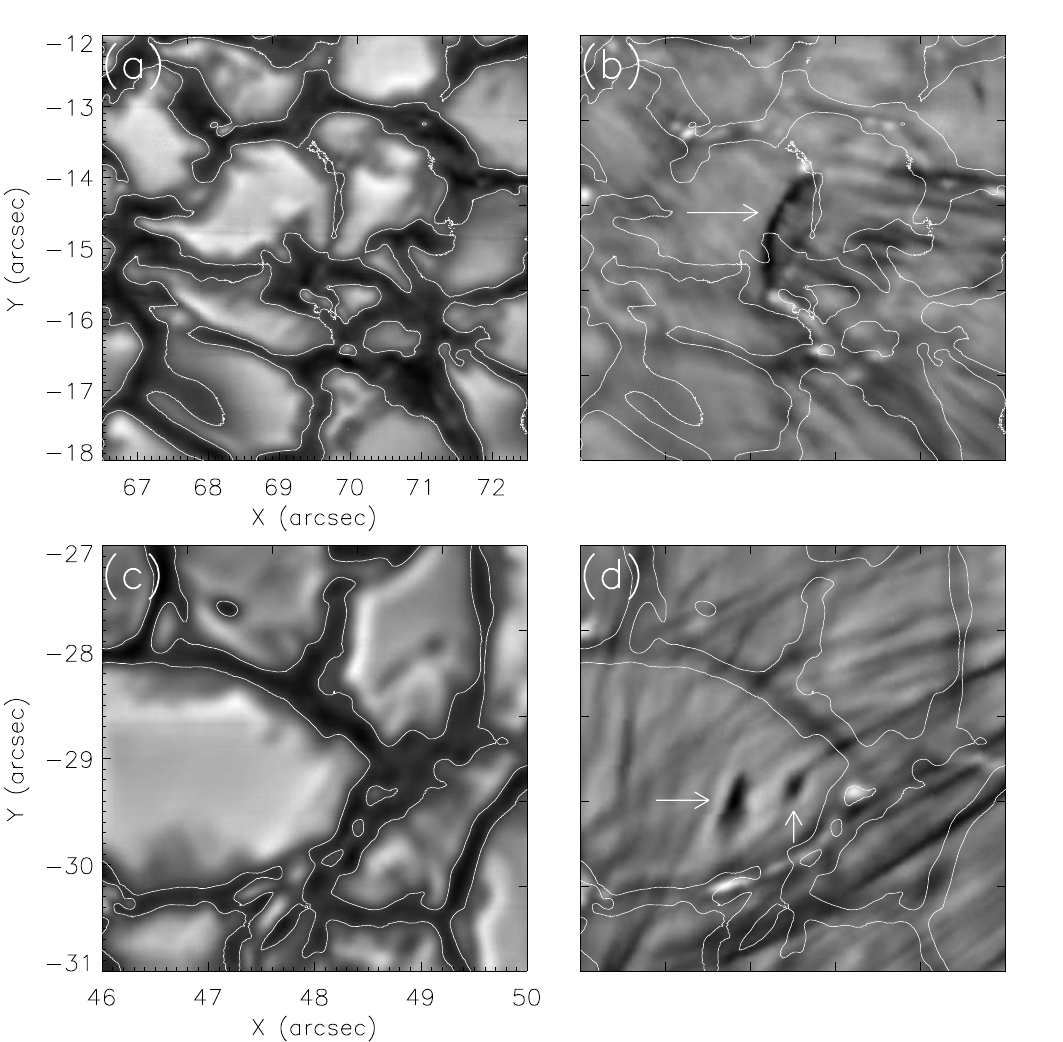}
	\caption{Panels (a, c): VBI $4503\,\mathrm{\AA}$ photospheric image. The white contours indicate intergranular lanes and are also marked in panel (b). Panels (b, d): Simultaneous VBI H$\alpha$ image. The white arrows point to three microfilaments, respectively. Panels (a–b) correspond to the event shown in Figure \ref{fig:fig2}(c), while panels (c–d) correspond to the event shown in Figure \ref{fig:fig3}.}
	\label{fig:figB1}
\end{figure*}

Since the VBI H$\alpha$ data are broadband imaging, small-scale H$\alpha$ dark features may resemble the morphology of photospheric intergranular lanes.
To resolve this ambiguity, we performed a strict spatial co-alignment between our H$\alpha$ broadband images and simultaneous photospheric observations from the VBI $4503\,\mathrm{\AA}$ (blue continuum) passband. A direct spatial comparison between the two passbands (shown in Figure \ref{fig:figB1}) demonstrates that the H$\alpha$ dark features do not spatially correspond to the intergranular lanes. Figures \ref{fig:figB1}(a–b) and (c–d) match the events presented in Figures \ref{fig:fig2}(c) and \ref{fig:fig3}, respectively.

\bibliography{references}{}
\bibliographystyle{aasjournalv7}

\end{document}